\documentclass[twocolumn, secnumarabic, amssymb, nobibnotes, aps]{revtex4-1}
\usepackage{amsmath,amstext,bm,appendix,graphicx,multirow}
\usepackage{color}
\usepackage{diagbox}
\usepackage{subfigure}
\usepackage{color}
\usepackage[bookmarks=false]{hyperref}
\hypersetup{colorlinks=true,citecolor=blue,linkcolor=blue,urlcolor=blue,pdfstartview=FitH,bookmarksopen=true}
\definecolor{green}{rgb}{0.8,0.98,0.83}
\pagecolor{white}
\date{\today}

\begin{document}
\title{Nonreciprocal Transmission and Entanglement in a cavity-magnomechanical system}

\author{Zhi-Bo Yang$^{1}$}%
\author{Jin-Song Liu$^{1}$}%
\author{Ai-Dong Zhu$^{1}$}%
\author{Hong-Yu Liu$^{1}$}%
\email{liuhongyu@ybu.edu.cn}
\author{Rong-Can Yang$^{2,3}$}%
\email{rcyang@fjnu.edu.cn}
\affiliation{$^{1}$Department of Physics, College of Science, Yanbian University, Yanji, Jilin 133002, China}
\affiliation{$^{2}$Fujian Provincial Key Laboratory of Quantum Manipulation and New Energy Materials, and College of Physics and Energy, Fujian Normal University, Fuzhou 350117, China}
\affiliation{$^{3}$Fujian Provincial Collaborative Innovation Center for Optoelectronic Semiconductors and Efficient Devices, Xiamen 361005, China}

\begin{abstract}
Quantum entanglement, a key element for quanum information is generated with a cavity magnomechanical system.
It comprises of two microwave cavities, a magnon mode and a vibrational mode, and the last two  elements come from a YIG sphere trapped in the second cavity. 
The two microwave cavities are connected by a superconducting transmission line, resulting in a linear coupling between them.
The magnon mode is driven by a strong microwave field and coupled to cavity photons via magnetic dipole interaction, and at the same time interacts with phonons via magnetostrictive interaction.  
By breaking symmetry of the configuration, we realize nonreciprocal photon transmission and one-way bipartite quantum entanglement. 
By using current experimental parameters for numerical simulation, it is hoped that our results may reveal a new strategy to built quantum resources for the realization of noise-tolerant quantum processors, chiral networks, and so on.
\end{abstract}
\maketitle
\section{introduction}
Although reciprocity is ubiquitous in nature, nonreciprocity promptes diverse applications such as chiral engineering, invisible sensing, and backaction-immune information processing~\cite{001,002,003}. 
So far, electromagnetic nonreciprocal transmission~\cite{004,005} has been demonstrated with various systems ranging from microwave~\cite{006,007,008,009}, terahertz~\cite{010}, optical~\cite{011,012} photons to $ x $-rays~\cite{013}. 
As a special type of nonreciprocity which only allows one-way transmission, unidirectional transmission of light is also revealed with some hybrid systems, such as atomic gases~\cite{014}, nonlinear devices~\cite{011,015,016,017}, moving media~\cite{018} and synthetic materials~\cite{019}. 
While in quantum regime, a recent scheme has been proposed, where nonreciprocal entangled states is prepared by using an optical diode with a high isolation rate~\cite{020}, making it possible to swift a single device between classical isolator and quantum diode, or to protect quantum resources from backscattering losses. 

Cavity spintronics~\cite{021,022,023,024,025,026,027,028,029,030,031,032,033,034,035} is an emerging and rapidly developing interdisciplinary that studies magnons strongly couple with microwave photons via magnetic dipole interaction. 
Such a hybrid system shows great application prospect in the field of quantum information processing, especially for quantum transducer~\cite{036,037,038,039,040,041,042} and quantum memory~\cite{043}. 
The reason is that magnetic materials have several distinguishing advantages such as long lifetime, high spin density and easy tunability. 
Furthermore, a collective excitation of spins in these materials \textbf{(}called magnon mode or Kittel mode\textbf{)} in magnetic materials can easily be coupled to a variety of other types of systems. 
Thus, cavity spintronics seems to be a potential candidate to study multiple quantum correlations, etc. 
By using cavity spintronics, in this manuscript, we propose a scheme to realize an optical diode with both quantum and classical characteristics, i.e., the implementation of both nonreciprocal microwave transmission and nonreciprocal bipartite quantum entanglement. 

In this manuscript, we present a proposal to carry out nonreciprocal microwave transmission and one-way bipartite entanglement by using an additional microwave cavity coupled to a cavity-magnon system to break symmetry of spatial inversion~\cite{009}. 
The two microwave cavities are linked by superconducting transmission lines~\cite{044} and one of them interacts with magnon mode of a ferrimagnetic yttrium iron garnet \textbf{(}YIG\textbf{)} sphere~\cite{045,046,047,048,049,021,024}. 
Simultaneously, the magnon mode is coupled to phonons due to vibration of YIG sphere induced by the magnetostrictive force~\cite{050,051,026}. 
In addition, magnons are driven by a strong microwave field at the first red sideband with respect to phonons because entanglement mainly survives with small thermal phonon occupancy.
The \textbf{(}anti-Stokes\textbf{)} process not only realizes phonon cooling, a prerequisite for observing quantum effects in the system~\cite{052}, but also also enhances the effective magnon-phonon coupling in order for the generation of magnomechanical entanglement. 
If both microwave cavities are resonant with the red sideband, then entanglement between the other two subsystems will be generated though some entanglement is very small. 
For the same driving power, different driving direction induces different effective magnetostrictive interaction, leading the generated subsystem entanglement to exhibit rare nonreciprocity and unidirectional invisibility. 
Furthermore, most of bipartite entanglement is robust against ambient temperature. 

\begin{figure}[b]
	\centering
	\includegraphics[width=1\linewidth,height=0.36\textheight]{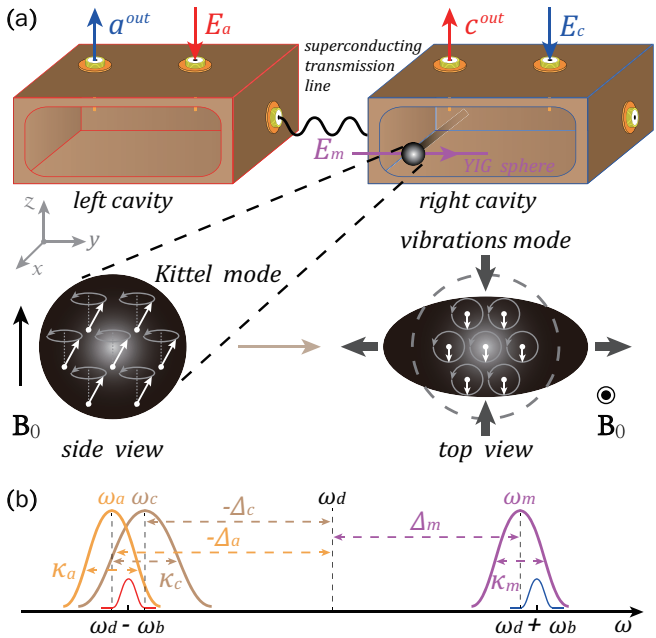}
	\hspace{0in}%
	\caption{(a) A schematic diagram of two-cavity magnomechanic system, consisting of two microwave cavities and a YIG sphere, which is placed in the right cavity. The YIG sphere, which is magnetized to saturation by a bias magnetic field $\textbf{B}_0$ aligned along the $ z $-direction, is mounted near the right cavity wall, where the magnetic field of the cavity mode is the strongest and polarized along $ x $-direction to excite the magnon mode in YIG. The magnon mode is driven by a microwave field along the $ y $-direction. Three magnetic fields are mutually perpendicular ar the site of the YIG sphere. In addition, the system will exhibit varying degrees of magnomechanical interaction when the left or right cavity is driven alone. (b) Mode frequencies and linewidths. The magnon mode \textbf{(}two MW cavity modes\textbf{)} with frequency $\omega_m$ \textbf{(}$\omega_a$ and $\omega_c$\textbf{)} is driven by a strong MW field at frequency $\omega_d$, and the mechanical motion of frequency $\omega_b$ scatters the driving photons onto two sidebands at $\omega_d\pm\omega_b$. If the magnon mode is resonant with the blue \textbf{(}anti-Stokes\textbf{)} sideband, and two cavity modes are resonant with the red \textbf{(}Stokes\textbf{)} sideband, all subsystems are prepared in entangled state. See text for more details.}
	\label{fig1}
\end{figure}

The manuscript is organized as follows. At first, we present a general model of the scheme, and then solve the system dynamics by means of the standard Langevin formalism with linearization treatment. Next, we illustrate nonreciprocity of microwave transmission and bipartite entanglement in the stationary state. Finally, we show how to measure generated entanglement and analyze the validity of our model.

\section{Model and equation of motion}
As schematically shown in Fig.~\ref{fig1} (a), we study a cavity magnomechanical system which consists of two coupled microwave cavities and one of them coupled to a YIG sphere.
The sphere is uniformly magnetized to saturation by a bias magnetic field with $\textbf{B}_0=B_0\textbf{e}_z$, where $B_0$ and $\textbf{e}_z$ represent magnetic amplitude and the unit vector in $z$ direction, respectively~\cite{023}. At the same time, the YIG sphere is also directly driven by a microwave field with driving strength $ E_m $ and frequency $\omega_d$. In addition, either the first cavity or the second one is driven by a microwave light beam with driving strength $ E_a $ or $ E_c $ with the same frequency $\omega_d$. The two cavities are linked by a superconducting transmission line~\cite{044} and the magnon mode couples to the second cavity mode and a vibrational mode~\cite{021} via the collective magnetic-dipole interaction and magnetostrictive interaction, respectively. The magnetostatic mode with finite wave number has a distinct frequency different from the Kittel mode so that the selective excitation may be implemented through the driving field wavelength and cavity mode selection. When all of the driving fields are included, the total Hamiltonian of the system can be written as follows 
\begin{eqnarray}\label{e001}
\mathcal{H}/\hbar&=&\omega_a\hat{a}^{\dag}\hat{a}+\omega_c\hat{c}^{\dag}\hat{c}+\omega_m\hat{m}^{\dag}\hat{m}+\omega_b\hat{b}^{\dag}\hat{b}+g_{ac}(\hat{a}^{\dag}\hat{c}+\hat{c}^{\dag}\hat{a})\nonumber\\
&&+g_{cm}(\hat{c}^{\dag}\hat{m}+\hat{m}^{\dag}\hat{c})+g_{mb}\hat{m}^{\dag}\hat{m}(\hat{b}^{\dag}+\hat{b})\nonumber\\
&&+iE_a(\hat{a}^{\dag}e^{-i\omega_dt}-\hat{a}e^{i\omega_dt})+iE_c(\hat{c}^{\dag}e^{-i\omega_dt}-\hat{c}e^{i\omega_dt})\nonumber\\
&&+iE_m(\hat{m}^{\dag}e^{-i\omega_dt}-\hat{m}e^{i\omega_dt}).
\end{eqnarray}
Here, $ \hat{a} $, $\hat{c}$, $\hat{m}$ and $\hat{b}$ \textbf{(}$ \hat{a}^{\dag} $, $ \hat{c}^{\dag} $, $ \hat{m}^{\dag} $ and $ \hat{b}^{\dag} $\textbf{)} are individually the annihilation \textbf{(}creation\textbf{)} operators of corresponding cavities, magnons and phonons with resonance frequency $ \omega_a $, $ \omega_c $, $ \omega_m $ and $ \omega_b $. The magnon frequency is determined by the external bias magnetic field $B_0$, i.e., $\omega_m=\gamma B_0$ with $\gamma/2\pi=2.8$ MHz/Oe being the gyromagnetic ratio. $g_{ac}$ is the photon coupling rate~\cite{053} and $g_{cm}$ represents the linear photon-magnon coupling strength which can be estimated by measuring the reflection spectrum of YIG sphere in right cavity and can be adjusted by varying the direction of the bias field or the position of the YIG sphere inside the right cavity~\cite{050}. The single-magnon magnomechanical coupling rate $ g_{mb} $ is typically small, but the magnomechanical interaction can be enhanced if magnons are strongly driven ~\cite{021,022,054,055}. The amplitude for each driving fields is $E_{i}=\sqrt{\kappa_i}\varepsilon_i$ with the effective strength $ \varepsilon_i=\sqrt{P_{i}/\hbar\omega_d} $ with the corresponding driving power and driving frequency being $ P_i $ and $ \omega_d $, respectively~\cite{025}. $\kappa_{a/c}$ and $\kappa_{m/b}$ separately represent the total loss rate of the first/second cavity and the magnon/phonon mode.

In the frame rotating at the driving frequency $\omega_d$, the quantum Langevin equations \textbf{(}QLEs\textbf{)} describing the system are given by
\begin{eqnarray}\label{e002}
\dot{\hat{a}}&=&-(i\Delta_a+\kappa_{a})\hat{a}-ig_{ac}\hat{c}+E_a+\sqrt{2\kappa_{a}}\hat{a}^{in},\nonumber\\
\dot{\hat{c}}&=&-(i\Delta_c+\kappa_{c})\hat{c}-ig_{ac}\hat{a}-ig_{cm}\hat{m}+E_c+\sqrt{2\kappa_{c}}\hat{c}^{in},\nonumber\\
\dot{\hat{m}}&=&-(i\Delta_m+\kappa_{m})\hat{m}-ig_{cm}\hat{c}-ig_{mb}\hat{m}(\hat{b}^{\dag}+\hat{b})+E_m\nonumber\\
&&+\sqrt{2\kappa_{m}}\hat{m}^{in},\nonumber\\
\dot{\hat{b}}&=&-(i\omega_b+\kappa_{b})\hat{b}-ig_{mb}\hat{m}^{\dag}\hat{m}+\sqrt{2\kappa_{b}}\hat{b}^{in},
\end{eqnarray}
where $\Delta_{j}=\omega_{j}-\omega_d$ and $\hat{a}^{in}$, $\hat{c}^{in}$, $\hat{m}^{in}$, $\hat{b}^{in}$ are input noise operators with zero mean value acting on the cavities, magnon and mechanical modes, respectively, which are characterized by the following correlation functions: $\langle \hat{a}^{in}(t)\hat{a}^{in\dag}(t^{'}) \rangle=\langle \hat{c}^{in}(t)\hat{c}^{in\dag}(t^{'}) \rangle=\langle \hat{m}^{in}(t)\hat{m}^{in\dag}(t^{'}) \rangle=\delta(t-t^{'})$, $\langle \hat{b}^{in}(t)\hat{b}^{in\dag}(t^{'}) \rangle=(n_b+1)\delta(t-t^{'})$, and $\langle \hat{b}^{in\dag}(t)\hat{b}^{in}(t^{'}) \rangle=n_b\delta(t-t^{'})$ where the equilibrium mean thermal phonon numbers $n_b=[\exp(\frac{\hbar\omega_b}{k_BT})-1]^{-1}$ with $ k_B $ the Boltzmann constant and $ T $ the ambient temperature.

Because the magnon mode and microwave photons are strongly driven, they all have large amplitude, i.e., $\arrowvert \langle a \rangle \arrowvert$, $\arrowvert \langle c \rangle \arrowvert$, $\arrowvert \langle m \rangle \arrowvert$ $\gg1$, which allows us to linearize the dynamics of the system around the steady-state values by writing any a operator as $\hat{o}=\langle o\rangle+\delta o$ \textbf{(}$o=a,c,m,b$\textbf{)} and neglecting second order fluctuation terms. Then, we obtain a set of differential equations for the mean values:
\begin{eqnarray}\label{e003}
\dot{\langle a \rangle}&=&-(i\Delta_a+\kappa_{a})\langle a \rangle-ig_{ac}\langle c \rangle+E_a,\nonumber\\
\dot{\langle c \rangle}&=&-(i\Delta_c+\kappa_{c})\langle c \rangle-ig_{ac}\langle a \rangle-ig_{cm}\langle m \rangle+E_c,\nonumber\\
\dot{\langle m \rangle}&=&-(i\tilde{\Delta}_m+\kappa_{m})\langle m \rangle-ig_{cm}\langle c \rangle+E_m,\nonumber\\
\dot{\langle b \rangle}&=&-(i\omega_b+\kappa_{b})\langle b \rangle-ig_{mb}\arrowvert \langle m \rangle \arrowvert^2,
\end{eqnarray}
and the linearized QLEs for the quantum fluctuations:
\begin{eqnarray}\label{e004}
\delta\dot{a}&=&-(i\Delta_a+\kappa_{a})\delta a-ig_{ac}\delta c+\sqrt{2\kappa_{a}}\hat{a}^{in},\nonumber\\
\delta\dot{c}&=&-(i\Delta_c+\kappa_{c})\delta c-ig_{ac}\delta a-ig_{cm}\delta m+\sqrt{2\kappa_{c}}\hat{c}^{in},\nonumber\\
\delta\dot{m}&=&-(i\tilde{\Delta}_m+\kappa_{m})\delta m-ig_{cm}\delta c-\frac{1}{2}G_{mb}(\delta b^{\dag}+\delta b)\nonumber\\
&&+\sqrt{2\kappa_{m}}\hat{m}^{in},\nonumber\\
\delta\dot{b}&=&-(i\omega_b+\kappa_{b})\delta b-\frac{1}{2}G_{mb}(\delta m^{\dag}-\delta m)+\sqrt{2\kappa_{b}}\hat{b}^{in},\nonumber\\
\end{eqnarray}
where $\tilde{\Delta}_m=\Delta_m+g_{mb}(\langle b \rangle+\langle b \rangle^*)$ is the effective detuning of the magnon mode including the frequency shift caused by the magnetostrictive interaction. $G_{mb}=i2g_{mb}\langle m \rangle$ is the effective magnomechanical coupling rate. If we consider the time to be $t\to \infty$ and the detunings to satisfy $\arrowvert\Delta_a\arrowvert$, $\arrowvert\Delta_c\arrowvert$, $\arrowvert \tilde{\Delta}_m \arrowvert$ $\gg$ $\kappa_{a}$, $\kappa_{c}$, $\kappa_{m}$, then $\langle a \rangle$, $\langle c \rangle$, $\langle m \rangle$ and $\langle b \rangle$ can be given by

\begin{eqnarray}\label{e007}
	\langle a \rangle&\simeq&i\frac{E_mg_{ac}g_{cm}-E_ag_{cm}^2-E_cg_{ac}\tilde{\Delta}_m+E_a\Delta_{c}\tilde{\Delta}_m }{g_{cm}^2\Delta_{a}+g_{ac}^2\tilde{\Delta}_m-\Delta_{a}\Delta_{c}\tilde{\Delta}_m },\nonumber\\
	\langle c \rangle&\simeq&-i\frac{E_mg_{cm}\Delta_{a}+E_ag_{ac}\tilde{\Delta}_m-E_c\Delta_{a}\tilde{\Delta}_m }{g_{cm}^2\Delta_{a}+g_{ac}^2\tilde{\Delta}_m-\Delta_{a}\Delta_{c}\tilde{\Delta}_m },\nonumber\\
	\langle m \rangle&\simeq&-i\frac{E_mg_{ac}^2-E_ag_{ac}g_{cm}+E_cg_{cm}\Delta_a-E_m\Delta_a\Delta_c}{g_{cm}^2\Delta_a+g_{ac}^2\tilde{\Delta}_m-\Delta_a\Delta_c\tilde{\Delta}_m},\nonumber\\
	\langle b \rangle&=&-ig_{mb}\arrowvert\langle m \rangle \arrowvert^2/(i\omega_b+\kappa_{b}).\nonumber\\
\end{eqnarray}
In what follows, we first show that the effect of the related parameters on the nonreciprocal transmission of a microwave field with a forward and backward driving-field input by solving the equations numerically, and then we show that the classical nonreciprocity can also be used to prepare nonreciprocal subsystem entangled states.

\section{Nonreciprocal transmission}
\begin{figure}
	\centering
	\includegraphics[width=0.9\linewidth,height=0.31\textheight]{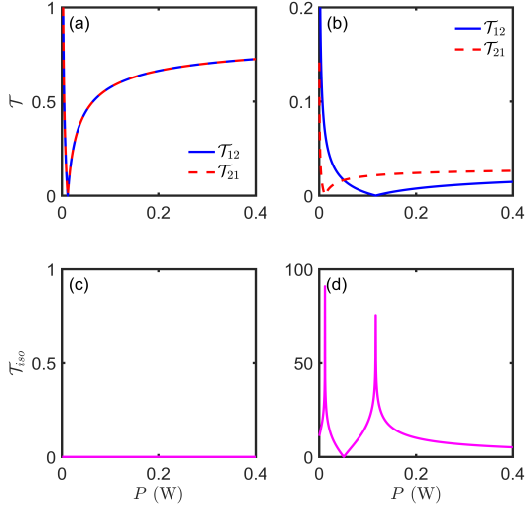}
	\hspace{0in}%
	\caption{(a) (b) Transmission coefficients $\mathcal{T}_{12}$ and $\mathcal{T}_{21}$ and (c) (d) transmission isolation ratio $T_{iso}$ versus driving power $P$ for (a) (c) $g_{ac}/\omega_b=1$ and (b) (d) $g_{ac}/\omega_b=0.32$, where $P_m=94.5$ mW. Other parameters are $\omega_a/2\pi=10$ GHz, $\omega_b/2\pi=10$ MHz, $\tilde{\Delta}_m=0.9\omega_b$, $\Delta_c=\Delta_a=-\omega_b$, $\kappa/2\pi=\kappa_{m}/2\pi=1$ MHz, and $g_{cm}/2\pi=3.2$ MHz which are mostly based on the latest experimental parameters~\cite{054,023}.}
	\label{fig2}
\end{figure}

In order to study the nonreciprocity of cavity output field, we let the first cavity mode driven \textbf{(}$E_a\neq0\quad$ \& $\quad E_c=0$\textbf{)} or the second cavity mode driven \textbf{(}$E_a=0\quad$ \& $\quad E_c\neq0$\textbf{)} by a classical field, and measure the output field of the other cavity~\cite{025}. For the sake of concision, we denote the first \textbf{(}second\textbf{)} case as the driving-field input from the forward \textbf{(}backward\textbf{)} direction in the following passages. Thus, the transmission coefficients for the two cases are separately defined as $ \mathcal{T}_{12}=\arrowvert c^{out}/\varepsilon_a \arrowvert = \arrowvert\sqrt{\kappa_{c}}\langle c \rangle/\varepsilon_a \arrowvert$ and $ \mathcal{T}_{21}=\arrowvert a^{out}/\varepsilon_c \arrowvert =\arrowvert\sqrt{\kappa_{a}}\langle a \rangle/\varepsilon_c \arrowvert$ with the average number of the first \textbf{(}second\textbf{)} cavity $\arrowvert \langle a(c) \rangle \arrowvert$ calculated with Eq.~(\ref{e007}). In addition, in order to depict the nonreciprocality, we define the isolation $ \mathcal{T}_{iso}=20\times log_{10}\arrowvert \mathcal{T}_{12}/\mathcal{T}_{21} \arrowvert $ which is given in decibels \textbf{(}dB\textbf{)}~\cite{009,025}. For the sake of simplicity, we set the power of the microwave source focused on the first or second cavity to be the same, i.e. $ P_a=P_c=P $, and the two cavities to be identical, i.e. $\kappa_a=\kappa_c=\kappa$. The transmission coefficient $ \mathcal{T}_{12/21} $ and the isolation $ \mathcal{T}_{iso} $ as a function of the driving power $P$ are plotted in Fig.~\ref{fig2} (a) and (c) with $g_{ac}/\omega_b=1$, and (b) and (d) with $g_{ac}/\omega_b=0.32$. The other parameters are chosen as $\omega_a/2\pi=10$ GHz, $\omega_b/2\pi=10$ MHz, $\tilde{\Delta}_m=0.9\omega_b$, $\Delta_c=\Delta_a=-\omega_b$, $\kappa/2\pi=\kappa_{m}/2\pi=1$ MHz, and $g_{cm}/2\pi=3.2$ MHz~\cite{054,023}.

From Eq.~(\ref{e007}), we find that when $g_{ac}=\omega_b$ the same transmission coefficient is observed for the different directions \textbf{(}forward or backward injection\textbf{)} of the driving-field input. We regard $g_{ac}=\omega_b$ as a impedance-matching condition~\cite{053} for the same microwave transmission. From Fig.~\ref{fig2} (a) with  $g_{ac}/\omega_b=1$, we can clearly see that the two transmission coefficients \textbf{(}$\mathcal{T}_{12}$ and $\mathcal{T}_{21}$\textbf{)} are always equal. However, if we decrease the coupling rate $g_{ac}$ between the two cavities, then the two coefficients \textbf{(}$\mathcal{T}_{12}$ and $\mathcal{T}_{21}$\textbf{)} will be distinctly different when $P$ is small. For example, $\mathcal{T}_{12}\simeq0.08,0$ and $\mathcal{T}_{21}\simeq0,0.02$ when $P\simeq10,110$ mW \textbf{(}see Fig.~\ref{fig2} (b)\textbf{)}. While the increase of $P$ will lead $\mathcal{T}_{12}$ and $\mathcal{T}_{21}$ to be closer and then to keep some difference. Therefore, if the impedance-matching condition is broken, the transmission for both two directions will be distinct. This is a typical nonreciprocal phenomenon, which corresponds to the classical nonreciprocal microwave transmission~\cite{009}. In addition, the degree of the nonreciprocal transmission of a microwave field can be described by the isolation as follows: $\mathcal{T}_{iso}=20\times log_{10}\arrowvert \mathcal{T}_{12}/\mathcal{T}_{21} \arrowvert$ given in decibels \textbf{(}dB\textbf{)}. The calculation results for $ \mathcal{T}_{iso}$, plotted as a function of driving power, is shown in Fig.~\ref{fig2} (c) and (d). Fig.~\ref{fig2} (c) shows that when the impedance-matching condition is met, the microwave transmission coefficients in the two directions are almost the same, i.e., $\mathcal{T}_{iso}=0$. However, Fig.~\ref{fig2} (d) gives us a clearer perspective for the differences between the transmission coefficients $\mathcal{T}_{12}$ and $\mathcal{T}_{21}$. At this time, the impedance-matching condition is broken, and the microwave transmission in the two directions has a larger transmission isolation ratio $\mathcal{T}_{iso}>70$ dB. It is worth noting that when $P=0$, $\mathcal{T}_{iso}\not=0$ in Fig.~\ref{fig2} (d), this is because the magnons mode is always driven by a microwave source with driving power $P=94.5$ mW to maintain a certain magnitude of magnomechanical coupling. 

\section{Definition of entanglement}
We thereby obtain the linearized QLEs for the quadratures $\delta X_o,\delta Y_o$ defined as $\delta X_o=(\delta o+\delta o^{\dag})/\sqrt{2},\delta Y_o=(\delta o-\delta o^{\dag})/i\sqrt{2}$, can be written as $\dot{\sigma}(t)=A\sigma(t)+n(t)$ with $\sigma(t)=[\delta X_a(t),\delta Y_a(t),\delta X_c(t),\delta Y_c(t),\delta X_m(t),\delta Y_m(t),\delta X_b(t),\\\delta Y_b(t)]^{T}$ and $n(t)=[\sqrt{2\kappa_{a}}\delta X_a^{in}(t),\sqrt{2\kappa_{a}}\delta Y_a^{in}(t),\sqrt{2\kappa_{c}}\\\delta X_c^{in}(t),\sqrt{2\kappa_{c}}\delta Y_c^{in}(t),\sqrt{2\kappa_{m}}\delta X_m^{in}(t),\sqrt{2\kappa_{m}}\delta Y_m^{in}(t),\\\sqrt{2\kappa_{b}}\delta X_b^{in}(t),\sqrt{2\kappa_{b}}\delta Y_b^{in}(t)]^T$ being the vectors for quantum fluctuations and noise, respectively. In addition, the drift matrix $A$ reads
\begin{eqnarray}\label{e006}
A=
\begin{bmatrix}
-\kappa_{a} & \Delta_a & 0 & g_{ac}& 0& 0& 0& 0\\ 
-\Delta_a &-\kappa_{a}&-g_{ac}&0& 0& 0& 0& 0\\
0&g_{ac}&-\kappa_c&\Delta_{c}&0&g_{cm}&0&0\\
-g_{ac}&0&-\Delta_{c}&-\kappa_c&-g_{cm}&0&0&0\\
0& 0& 0& g_{cm}&-\kappa_m& \tilde{\Delta}_m& -G_{mb}& 0\\
0& 0& -g_{cm}& 0& -\tilde{\Delta}_m&-\kappa_m& 0& 0\\
0& 0& 0& 0& 0& 0&-\kappa_b& \omega_b\\
0& 0& 0& 0& 0& G_{mb}& \omega_b&-\kappa_b\\
\end{bmatrix}.\nonumber\\
\end{eqnarray}
In this case, when the forward input direction of the driving field is taken into account \textbf{(}i.e., $E_a\not=0$ and $E_c=0$\textbf{)}, magnomechanical coupling can be written as follows
\begin{eqnarray}\label{e008}
G_{mb,12}&=&\frac{2g_{mb}(E_mg_{ac}^2-E_ag_{ac}g_{cm}-E_m\Delta_a\Delta_c)}{g_{cm}^2\Delta_a+g_{ac}^2\tilde{\Delta}_m-\Delta_a\Delta_c\tilde{\Delta}_m}.
\end{eqnarray}
However, when the backward input direction of the driving field is taken into account \textbf{(}i.e., $E_a=0$ and $E_c\not=0$\textbf{)}, magnomechanical coupling changes to 
\begin{eqnarray}\label{e009}
G_{mb,21}&=&\frac{2g_{mb}(E_mg_{ac}^2+E_cg_{cm}\Delta_a-E_m\Delta_a\Delta_c)}{g_{cm}^2\Delta_a+g_{ac}^2\tilde{\Delta}_m-\Delta_a\Delta_c\tilde{\Delta}_m}.
\end{eqnarray}
Since we are using linearized QLEs, the Gaussian nature of the input states will be preserved during the time evolution for the system. So, the quantum fluctuation is in a continuous four-mode Gaussian state which can be completely characterized by a $ 8\times 8 $ covariance matrix \textbf{(}CM\textbf{)} $ V $ in the phase space $ 2V_{ij}(t,t^{'})=\langle \sigma_i(t)\sigma_j(t^{'})+\sigma_j(t^{'})\sigma_i(t) \rangle $, \textbf{(}$ i,j=1,\cdots,8 $\textbf{)}. Then the vector can be obtained straightforwardly by solving the Lyapunov equation
\begin{eqnarray}\label{e010}
AV+VA^T=-D
\end{eqnarray}
with $D=$ diag $[\kappa_a,\kappa_a,\kappa_c,\kappa_c,\kappa_m,\kappa_m,(2n_b+1)\kappa_b,(2n_b+1)\kappa_b] $ defind through $ 2D_{ij}\delta (t-t^{'})=\langle n_i(t)n_j(t^{'})+n_j(t^{'})n_i(t) \rangle $. In this manuscript, we use logarithmic negativity~\cite{058,059} to quantify the degree of the quantum entanglement for the four-mode Gaussian state, which is defined as $E_N\equiv\max[0,-2\ln\nu^-]$, where $\nu^-=\min $ eig $ \arrowvert \oplus_{j=1}^{2}(-\sigma_y)PVP\arrowvert$ with $\sigma_y$ and $P=\sigma_z\oplus1$ are respectively $y$-Pauli matrix and  the matrix that realizes partial transposition at the CM level~\cite{058,060}.
\begin{figure}[b]
	\centering
	\includegraphics[width=0.9\linewidth,height=0.31\textheight]{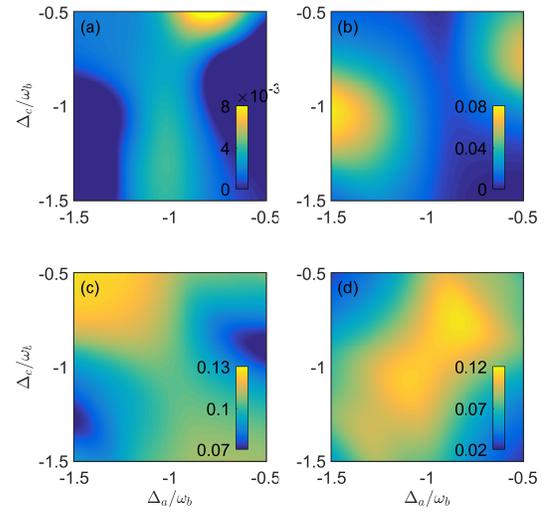}
	\hspace{0in}%
	\caption{(a) $E_{ac}$, (b) $E_{cm}$, (c) $E_{mb}$ and (d) $E_{ab}$ versus detunings $\Delta_a$ and $\Delta_{c}$. We take $G_{mb}/2\pi=2.5$ MHz, $\kappa_{b}/2\pi= 100$ Hz, $T=20$ mK and the other parameters are same as Fig.~\ref{fig2}.}
	\label{fig3}
\end{figure}

\section{Nonreciprocal entanglement}
The foremost task of studying entanglement properties between any two subsystems in such a hybrid system is to find the optimal effective interaction among modes, i.e., to find optimal frequency detuning that can generate subsystem entanglement~\cite{021}. In Fig.~\ref{fig3}, we show four types of subsystem entanglement \textbf{(}$E_{ac}$, $E_{cm}$, $E_{mb}$ and $E_{ab}$\textbf{)} as the function of cavity detunings $\Delta_{a}$ and $\Delta_{c}$, where $E_{ac}$, $E_{cm}$, $E_{mb}$ and $E_{ab}$ denote the cavity-cavity entanglement, cavity-magnon entanglement, magnon-phonon entanglement, and cavity-phonon distant entanglement, respectively. In addition, we choose $G_{mb}/2\pi=2.5$ MHz, $\kappa_{a}=\kappa_{c}=\kappa_{m}$, $\kappa_{b}/2\pi= 100$ Hz, $T=20$ mK and the other parameters are chosen as the same as that in Fig.~\ref{fig2}. Furthermore, we also set $\tilde{\Delta}_m\simeq\omega_b$ which imply that magnon mode is in the blue sideband with respect to the first cavity mode, which corresponds to the anti-Stokes process, i.e., significantly cooling the phonon mode. Thus, the elimination of the main obstacle for observing entanglement is obtained~\cite{021}. It is noted that all results are satisfying with the condition of the steady state guaranteed by the negative eigenvalues \textbf{(}real parts\textbf{)} of the drift matrix $ A $. From Fig.~\ref{fig3}, it is shown that a parameter regime exists, i.e., $\Delta_a=\Delta_c=-\omega_b$, where the entanglement within any two subsystems occurs \textbf{(}see Fig.~\ref{fig1} (b)\textbf{)}. This is similar to the realization of entanglement between magnon modes in a magnomechanic system with two YIG spheres proposed in Ref.~\cite{061}. In order to obtain the entanglement between any two subsystems and keep the system stable at the same time, the three coupling rates $g_{ac}$, $g_{cm}$ and $G_{mb}$ should be on the same order of magnitude and chosen as a separate and moderate value. Based on it, we choose  $g_{ac}g_{cm}G_{mb}\ll\arrowvert \Delta_{a}\Delta_{c}\tilde{\Delta}_m \arrowvert\simeq\omega_b^3$ and $ G_{mb}=2.5 $ MHz  which corresponds to $E_m\simeq G_{mb}\omega_b^3/2g_{mb}(\omega_b^2-g_{ac}^2)\simeq4.2\times10^{13}$ Hz and $P_m\simeq1.85$ mW, when $E_a=E_c=0$ in Fig.~\ref{fig3}. All entanglement in the system comes from the magnetostrictive interaction between the magnon and the phonon~\cite{021}. Next, we will show the realization of nonreciprocal subsystem entanglement by controlling the classical nonreciprocal transmission, i.e., controlling the effective magnomechanical interaction. 
\begin{figure}[b]
	\centering
	\includegraphics[width=0.9\linewidth,height=0.31\textheight]{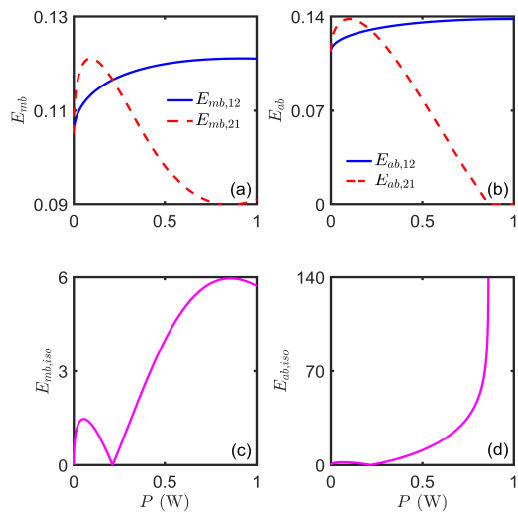}
	\hspace{0in}%
	\caption{(a) $E_{mb}$, (b) $E_{ab}$, (c) $E_{mb,iso}$, and (d) $E_{ab,iso}$ versus driving power $P$. We take $g_{mb}/2\pi=0.3$ Hz, $\kappa_{b}/2\pi= 100$ Hz, $T=20$ mK and the other parameters are same as Fig.~\ref{fig2}.}
	\label{fig4}
\end{figure}

For the sake of brevity, we mainly focus on the entanglement $E_{ab}$ and $E_{mb}$ due to they being larger than the other types of entanglement. In Fig.~\ref{fig4} (a) (b), we show $E_{mb}$ and $E_{ab}$ as the function of driving power $P$ on the cavity $a$ or $c$, where $E_{ij,12}$ \textbf{(}$E_{ij,21}$\textbf{)} represents the entanglement between mode $i$ and mode $ j $ when the direction of driving microwave source is forward \textbf{(}backward\textbf{)}. Additionally, $g_{mb}/2\pi=0.3$ Hz, $\kappa_{a}=\kappa_{c}=\kappa_{m}$ and $T=20$ mK are set and the other parameters are chosen as the same as that in Fig.~\ref{fig2}. Due to the existence of magnon-phonon nonlinear coupling \textbf{(}magnetostrictive force\textbf{)}, the entanglement of the magnomechanical subsystem is generated. And because there is a direct or indirect linear state-swap interaction between the two microwave cavities and the magnon mode, the resulting magnomechanical entanglement is distributed to other subsystems. Therefore, controlling the optimal effective magnomechanical interaction will be an effective operation to control whether entanglement exists in each subsystem. From Fig.~\ref{fig4} (a) we can see that the magnomechanical subsystem entanglement in the two directions shows very different trends with the increase of the driving power. Therefore, entanglement of other subsystems will be affected accordingly \textbf{(}see Fig.~\ref{fig4} (d)\textbf{)}. The results in Fig.~\ref{fig4} (a) (b) demonstrate that the entanglement of multiple subsystems in a hybrid system can be prepared in a highly asymmetric way. This is in good agreement with our previous expectations. This is a clear signature of quantum nonreciprocity, which is fundamentally different from that in classical devices~\cite{062,063} showing only nonreciprocal transmission rates.

\begin{figure}[b]
	\centering
	\includegraphics[width=0.9\linewidth,height=0.31\textheight]{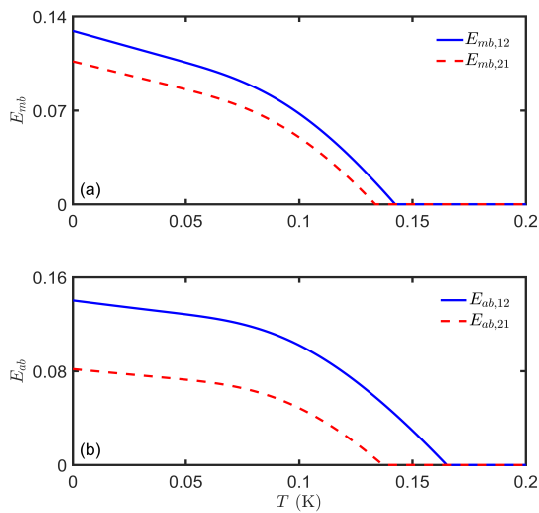}
	\hspace{0in}%
	\caption{(a) $E_{mb}$ and (b) $E_{ab}$ versus ambient temperature $T$. We take $P=0.5$ W, $g_{mb}/2\pi=0.3$ Hz, $\kappa_{b}/2\pi= 100$ Hz and the other parameters are same as Fig.~\ref{fig2}.}
	\label{fig5}
\end{figure}
Next, the difference between subsystem entanglement with forward driving direction and backward driving direction is extracted in the decibel scale \textbf{(}defined as $ 20\times log_{10}\arrowvert E_{ij,12}/E_{ij,21} \arrowvert $, similar to $\mathcal{T}_{iso}$~\cite{003}\textbf{)}, and we take its value as the entanglement isolation ratio $ E_{ij,iso} $ \textbf{(}units of dB\textbf{)}. For the case of reciprocal subsystem entanglement, we have $E_{ij,12}/E_{ij,21}=1$ and $ E_{ij,iso}=0 $. A nonzero $ E_{ij,iso} $ presents nonreciprocal entanglement and the greater the value of $ E_{ij,iso} $, the higher is the degree of the nonreciprocal entanglement. The calculation results for $ E_{ij,iso} $, plotted as a function of the driving power $P$, are shown in Fig.~\ref{fig4} (c) (d), which gives us a clearer perspective for the differences between the entanglements $ E_{ij,12} $ and $ E_{ij,21} $. Notice the cutoff in Fig.~\ref{fig4} (d). The position of the cutoff corresponds to the position of $E_{ab,21}=0$ in Fig.~\ref{fig4} (b), which shows the unidirectional invisibility of subsystem entanglement. In the multilayer microwave integrated quantum circuit, we can further design superconducting transmission lines and interconnects based on these existing technologies to provide the large range of necessary couplings and to minimize any parasitic losses~\cite{044}. In fact, we can find from Eqs.~(\ref{e008}--\ref{e009}) that when $g_{ac}=\omega_b$, the system reaches the impedance matching condition~\cite{025,053}, i.e., $G_{mb,12}=G_{mb,21}$. This realizes the switch from nonreciprocal to reciprocal of subsystem entangled state and shows the potential advantage of our solution as a tunable quantum diode.

Fig.~\ref{fig5}, $E_{mb}$ and $E_{ab}$ as a function of ambient temperature $T$ for driving power $P=1$ W, shows that the generated subsystem entanglements is robust to ambient temperature and survives up to $\sim100$ mK, below which the average phonon number is always smaller than 1, showing that mechanical cooling is, thus, a precondition for observing quantum entanglement in the system~\cite{021}. Compared with the scheme that a strong squeezed vacuum field proposed in Ref.~\cite{064} is used to generate entanglement between magnon modes, in our scheme due to the inherent low frequency of phonon modes, this robustness is generally weak.

\section{Discussion and conclusion}
Lastly, we discuss how to detect the entanglement and verify the effectiveness of two-cavity magnomechanic system. The generated subsystem entanglements can be detected by measuring the CM of two cavity output fields~\cite{021}. Such measurement in the microwave domain has been realized in the experiments~\cite{065,066}. In addition, for a 0.5-mm-diameter YIG sphere, the number of spins $N\simeq2.8\times10^{17}$, and $P_m=189$ mW corresponds to $E_{m}\simeq3\times10^{14}$ Hz, and $\arrowvert \langle m \rangle \arrowvert\simeq5.9\times10^{6}$, leading to $\langle m^{\dag}m \rangle\simeq3.5\times10^{13}\ll5N=1.4\times10^{18}$ which is well fulfilled. It is worth noting that two cavity modes is resonant with the red-sideband \textbf{(}$\Delta_{a}=\Delta_{c}=-\omega_b$\textbf{)} results in a higher magnon excitation number in the stable state in the presence of $ E_a $ or $ E_c $, where the magnon number has a simpler form $ \langle m^{\dag}m \rangle\simeq [E_m(\omega_b^2-g_{ac}^2)+E_ag_{ac}g_{cm}]^2/\omega_b^6$ or $ \langle m^{\dag}m \rangle\simeq [E_m(\omega_b^2-g_{ac}^2)+E_c\omega_bg_{cm}]^2/\omega_b^6$. Therefore, under the premise that the magnon mode is continuously driven by the classical microwave field with driving power $ P_m=94.5 $ mW, $ P=1 $ W corresponds to $ \langle m^{\dag}m \rangle_{12}\simeq1.4\times10^{15} \ll5N$ \textbf{(}$E_a\not=0$, $E_c=0$\textbf{)} and $ \langle m^{\dag}m \rangle_{21}\simeq3.47\times10^{15} \ll5N$ ($E_c\not=0$, $E_a=0$), respectively. In order to keep the Kerr effect negligible, $ K\arrowvert \langle m \rangle \arrowvert^3\ll \sum_{i=a,c,m} E_i $ must hold~\cite{021}. Kerr coefficient $ K $ is inversely proportional to the volume of the sphere. In this manuscript, we use a 5-mm-diameter YIG sphere, $K/2\pi\simeq8\times10^{-10}$ Hz which corresponds to $K\arrowvert \langle m \rangle \arrowvert^3\simeq4.2\times10^{13}$ Hz $\ll E_m+E_a\simeq1.3\times10^{15}$ \textbf{(}$P_m=189$ mW, $P=2$ W, and $E_c=0$\textbf{)} and $K\arrowvert \langle m \rangle \arrowvert^3\simeq1.64\times10^{14}$ Hz $\ll E_m+E_c\simeq1.3\times10^{15}$ \textbf{(}$P_m=189$ mW, $P=2$ W, and $E_a=0$\textbf{)}. This implying that the nonlinear effects are negligible and the linearization treatment of the model is a good approximation.

In summary, we show how to use an asymmetric cavity magnomechanic system to produce classical nonreciprocal transmission, and extend this nonreciprocity to quantum states, thereby generating nonreciprocal subsystem entangled states, rather than by means of nonreciprocal devices~\cite{062}. The introduction of an additional cavity mode successfully breaks the symmetry of spatial inversion. Our work opens up a range of exciting opportunities for quantum information processing, networking and metrology by exploiting the power of quantum nonreciprocity. The ability to manipulate quantum states or nonclassical correlations in a nonreciprocal way sheds new lights on chiral quantum engineering and can stimulate more works on achieving and operating quantum nonreciprocal devices, such as directional quantum squeezing~\cite{067,068}, backaction-immune quantum sensing~\cite{069,070}, and quantum chiral coupling of cavity optomechanics devices to superconducting qubits or atomic spins~\cite{021,071,072}.

\section{Acknowledgments}
This work is supported by the  Science and Technology project of Jilin Provincial Education Department of China during the 13th Five-Year Plan Period \textbf{(}Grant No. JJKH20200510KJ\textbf{)} and the Fujian Natural Science Foundation \textbf{(}Grant No. 2018J01661 and No. 2019J01431\textbf{)}.


\begin{thebibliography}{999}
\bibitem{055}
	For a 250-$\mu$m diameter YIG sphere, the corresponding $g_{mb}/2\pi\simeq0.2$ Hz~\cite{021}. In this manuscript, however, in order to avoid undesired Kerr nonlinearity, we choose a YIG sphere with a larger volume (5-mm diameter), which leads to a smaller single-magnon magnomechanical coupling $ g_{mb} $. Nevertheless, when any cavity field and magnon mode are strongly driven at the same time, we can still obtain a strong $ G_{mb} $ under the premise that the Kerr effect is negligible.
\bibitem{056}
	The external loss rate of the cavity field comes from the connection of the superconducting transmission line and the input and output ports on the cavity. For the port-induced decay rates $\kappa_{a}^{ex}$ and $\kappa_{c}^{ex}$, their magnitudes can be tuned by adjusting the lengths of the port pins inside the cavity and when the YIG sphere is inserted into the cavity, both the cavity frequency and intrinsic loss rate shift slightly at different displacements of the YIG sphere~\cite{035}. However, the loss rate of the magnon comes from the surface roughness as well as the impurities and defects in the YIG sphere~\cite{049,032}.
\end{thebibliography}
\end{document}